\documentclass[final,5p,times,twocolumn]{elsarticle}
\usepackage{enumerate}
\usepackage{amsmath}
\usepackage{amsfonts}
\usepackage{amssymb}
\usepackage[utf8]{inputenc}
\usepackage[T1]{fontenc}
\usepackage{mathtools}
\usepackage{wasysym}
\usepackage{accents}
\usepackage{graphicx}
\usepackage[svgnames]{xcolor}
\usepackage[colorlinks=true,citecolor=blue]{hyperref}
\usepackage{url}
\usepackage{amsmath}
\DeclareMathOperator\arctanh{arctanh}
\DeclareMathOperator\sech{sech}

\usepackage{orcidlink}

\newcommand\Eo{E_{\circ}}
\newcommand\Po{P_{\circ}}
\newcommand\No{N_{\circ}}

\journal{Physics Letters B}

\begin{document}
\begin{frontmatter}

\author[JP2]{Iarley P. Lobo\,\orcidlink{0000-0002-1055-407X}}
\ead{lobofisica@gmail.com}

\affiliation[JP2]{organization={Department of Chemistry and Physics, Federal University of Para\'iba},
            addressline={Rodovia BR 079 - Km 12}, 
            city={Areia},
            postcode={58397-000}, 
            state={PB},
            country={Brazil}}

\title{Time Delay in $\kappa$-Anti-de Sitter Spacetime}

\begin{abstract}
Based on deformed translations in the $\kappa$-anti-de Sitter algebra, we derive a delay in the time of detection between a soft and a hard photon, which are simultaneously emitted at a distant event, to first order in the quantum gravity parameter. In the basis analyzed, the trajectories are undeformed, and the effect depends exclusively on the symmetry properties of the quantum algebra. The time delay depends linearly on the energy of the hard particle and has a sinusoidal dependence on the redshift of the source.
\end{abstract}


\end{frontmatter}


\section{Introduction}

The quantization of gravity is one of the most challenging problems in science and has been examined by generations of physicists and mathematicians over the past decades. Despite some success in constructing theories that capture different aspects of the expected nature of quantum spacetime or the quantum nature of the gravitational interaction, such as canonical quantum gravity \cite{Kiefer:2004xyv}, string theory \cite{Polchinski:1998rq,Polchinski:1998rr}, loop quantum gravity \cite{Ashtekar:2021kfp}, causal dynamical triangulation \cite{Loll:2019rdj}, and asymptotic safety \cite{Eichhorn:2018yfc}, no experimental verification that validates any of these approaches has been found yet.

A common result in many of these theories is that in the asymptotic limit approaching the low gravity regime or low energy scale, the physical observables are given by perturbations of the equations of special/general relativity or quantum mechanics. Furthermore, such corrections are suppressed by a scale that is usually proportional to Planck units. As this regime is so far-reaching when we consider typical energies of elementary particles in accelerators, the quantization of the gravitational interaction has been assumed as a purely mathematical and philosophical curiosity for many years. However, with the development of the field of astroparticle physics, this picture has changed. We are now able to detect particles with higher energies, and the distances over which particles propagate are much longer than in accelerators, amplifying the tiny Planckian corrections \cite{Amelino-Camelia:2008aez,Addazi:2021xuf}. These new inputs, along with improvements in detection precision by observatories, have led to the emergence of the field of quantum gravity phenomenology by the end of the 20th century. This field has achieved Planck scale sensitivity in constraining quantum gravity-inspired corrections in recent years \cite{fermigbmlat:2009nfe} and presents many appealing avenues to explore in the coming years \cite{AlvesBatista:2023wqm}.

The most explored avenue in this direction is the phenomenon called in-vacuo dispersion \cite{Addazi:2021xuf}, which is the dependence of massless particles' speed on their energy, and is a prediction of several approaches to quantum gravity \cite{Amelino-Camelia:2008aez,Gambini:1998it,Amelino-Camelia:2016gfx} and other forms of Lorentz Invariance Violation (LIV) scenarios, like extensions of the standard model theories \cite{Kostelecky:2003fs}. The standard assumption in the derivation of in-vacuo dispersion bounds is the Jacob-Piran formula \cite{Jacob:2008bw} (which amended the result of \cite{Ellis:2002in}), that gives the dependence on particles' energies and propagation distance for the time delay from the arrival of massless particles when they are simultaneously emitted from a given source.

As was shown in \cite{Amelino-Camelia:2012vzf,Rosati:2015pga}, the main assumptions of the Jacob-Piran formula \cite{Jacob:2008bw} are the minimal coupling between the particles' momentum and local violation of Poincaré symmetry (rotations, boosts, and translations). The relaxation of these hypotheses has been explored in some recent works on curvature-triggered effects \cite{Amelino-Camelia:2020bvx,Amelino-Camelia:2022pja,Amelino-Camelia:2023srg} and other ansatze for defining such delays \cite{Pfeifer:2018pty}. For instance, if instead of violation, one incorporates in the derivation a {\it deformation} of local Poincaré symmetry that preserves the quantum gravity departures of relativistic equations, one finds corrections to the Jacob-Piran formula that could better fit the data from astrophysical events, as shown in the preliminary work \cite{Bolmont:2022yad}. This deformation has been initially shown for the special case of assuming that the spacetime has a Planck-scale-deformed de Sitter structure, where a deformation of the de Sitter algebra has been the background of the derivation. Besides presenting deformed worldlines (that coincide with Jacob-Piran's ansatz), it also presents deformed translation transformations, which govern the communication between the distant frames given by the emission and detection events. We should stress that calculating such effects in a Deformed Special Relativity (DSR) scenario \cite{Amelino-Camelia:2000stu} using modified translations (following the lessons of the Principle of Relative Locality \cite{Amelino-Camelia:2011lvm}) is of utmost importance to avoid non-local pathologies pointed out by different authors \cite{Schutzhold:2003yp,Hossenfelder:2010tm}, whose solutions were discussed in \cite{Amelino-Camelia:2011uwb}.

Although the de Sitter case has been extensively explored in the literature, deformations of the anti-de Sitter (AdS) algebra from the point of view of time delays have been neglected, despite the dedicated theoretical work that has been devoted to rigorously formulating the features of a deformation of the anti-de Sitter algebra \cite{Ballesteros:2016bml}, in which not only the full set of isometries are explored, but also the co-product structure in AdS space for multiparticle systems and its Carrollian and Galilean limits \cite{Ballesteros:2021dob}.

In this letter, we fill this gap and derive a time delay formula that depends on the energy of a hard (high energy) massless particle simultaneously emitted with a soft one (low energy) in a given spacetime event, following the recipe discussed in \cite{Amelino-Camelia:2012vzf}, but applied to the deformation of the anti-de Sitter algebra described in \cite{Ballesteros:2016bml}, at first order in the quantum gravity scale.

The letter is organized as follows. In section \ref{sec:class-ads}, we revise the classical mechanics of anti-de Sitter spacetime by deriving the conserved charges from the Killing equations in cosmological coordinates and the associated Casimir that is a mass shell. In section \ref{sec:q-ads}, we find the dependence of the symmetry generators and Casimir charge of the $\kappa$-AdS algebra in terms of the undeformed ones. In section \ref{sec:trans-conf}, we express the translation generators in conformal coordinates, which are easier to handle computationally. In section \ref{sec:k-time-delay}, we use the deformed translations to derive the time delay in the $\kappa$-AdS spacetime from the basis rigorously explored in \cite{Ballesteros:2016bml}. We draw our final remarks in section \ref{sec:conc}. We assume natural units, such that $\hbar = c = 1$.



\section{Undeformed Anti-de Sitter Kinematics}\label{sec:class-ads}

Before we proceed to the derivation of the deformed kinematics of particles in anti-de Sitter geometry, we revisit its simpler, undeformed counterpart. In this section, we primarily rely on purely algebraic tools to derive our results; however, our starting point is the derivation of symmetry generators in AdS space. As is usually the case in similar analyses, we shall rely on the $1+1$ dimensional example. The generators of symmetries are conserved charges; therefore, they should be found from the Killing vectors in a given spacetime. For AdS space, we consider the metric given by the following line element of FLRW type:
\begin{equation}\label{line-element}
    ds^2=dt^2-\cos^2(\lambda t)dx^2=g_{\mu\nu}dx^{\mu}dx^{\nu}\, ,
\end{equation}
where the parameter $\lambda$ is related to the anti-de Sitter curvature as $\Lambda=-\lambda^2$. We solve the Killing equation $\nabla_{\mu}K_{\nu}+\nabla_{\nu}K_{\mu}=0$ for the Killing vector $K_{\mu}$, which in $1+1$ dimensions presents only three solutions, responsible for describing time translation $K^{\mu}_{(0)}$, space translation $K^{\mu}_{(1)}$, and spacetime boost $K^{\mu}_{(2)}$. The energy $E_0$, momentum $P_0$, and boost generator $N$ are conserved charges, given by
\begin{eqnarray}
E_{\circ}&=&K^{\mu}_{(0)}p_{\mu}=\cosh{(\lambda x)}p_0+\sinh{(\lambda x)}\tan{(\lambda t)} p_1\, ,\label{e-circ}\\
P_{\circ}&=&K^{\mu}_{(1)}p_{\mu}=p_1\, ,\label{p-circ}\\
N_{\circ}&=&K^{\mu}_{(2)}p_{\mu}=\frac{\sinh{(\lambda x)}}{\lambda}p_0+\cosh{(\lambda x)}\frac{\tan{(\lambda t)}}{\lambda}p_1\, ,\label{n-circ}
\end{eqnarray}
where we use the subscript ``$\circ$'' in these expressions to stress that we are considering the undeformed anti-de Sitter generators.

Considering that the generalized coordinates $(t,x,p_0,p_1)$ are canonical, i.e.,
\begin{equation}\label{can-brack}
\{t,p_0\}=1=\{x,p_1\}\, ,
\end{equation}
and zero otherwise, we verify that the above charges define a Lie algebra as follows:
\begin{eqnarray}
    \{E_{\circ},P_{\circ}\}&=&\lambda^2 N_{\circ}=-\Lambda N_{\circ}\, ,\label{poisson-ec-pc}\\
    \{N_{\circ},P_{\circ}\}&=&E_{\circ}\, ,\label{poisson-nc-pc}\\
    \{N_{\circ},E_{\circ}\}&=&P_{\circ}\, ,\label{poisson-nc-ec}
\end{eqnarray}
where the Poisson brackets of two phase space quantities $F$ and $G$ are 
\begin{equation}
\{F,G\}=\frac{\partial F}{\partial x^{\mu}}\frac{\partial G}{\partial p_{\mu}}-\frac{\partial F}{\partial p_{\mu}}\frac{\partial G}{\partial x^{\mu}} \, . 
\end{equation}

This is the known algebra of anti-de Sitter generators in $1+1$ dimensions, $\text{AdS}_{\Lambda}$ \cite{Ballesteros:2016bml}. The Casimir operator related to this algebra is given by Eq.(2.2) of \cite{Ballesteros:2016bml} and in cosmological coordinates reads:
\begin{equation}\label{casimir}
    C_{\circ}=E_{\circ}^2-P_{\circ}^2-\lambda^2 N_{\circ}^2=p_{0}^2-\sec^2{(\lambda t)} p_1^2\, ,
\end{equation}
which is simply the norm $g^{\mu\nu}p_{\mu}p_{\nu}$ using the metric defined by the line element \eqref{line-element}.

\subsection{Undeformed Trajectories}\label{subsec:class-traj}

The trajectories can be easily found using the Casimir \eqref{casimir} as the generator of the parametrization evolution (here ``dot'' means derivative with respect to an affine parameter):
\begin{eqnarray}
\dot{t}&=&\{t,C_{\circ}\}=2p_0\, ,\label{t-dot}\\
\dot{x}&=&\{x,C_{\circ}\}=-2p_1\sec^2{(\lambda t)}\, ,\label{x-dot}\\
\dot{p_0}&=&\{p_0,C_{\circ}\}=2\lambda p_1^2\sec^2{(\lambda t)}\tan^2{(\lambda t)}\, ,\\
\dot{p_1}&=&\{p_1,C_{\circ}\}=0\, .
\end{eqnarray}

To derive an equation of the form $x(t)$, we need to express the equation $dx/dt=\dot{x}/\dot{t}$ using Eqs. \eqref{t-dot} and \eqref{x-dot}. This gives an equation that depends on $t$ and $p_0$. However, we can also express $p_0$ as a function of $t$ for the trajectories of particles with mass $m$, i.e., those for which $C_{\circ}=m^2$ in \eqref{casimir}. This gives:
\begin{equation}
    \frac{dx}{dt}=-\frac{p_1\sec^2{(\lambda t)}}{\sqrt{m^2+\sec^2{(\lambda t)}p_1^2}}\, .\label{dxdt}
\end{equation}

This equation can be solved analytically for any mass $m$. For the massless case $m\rightarrow 0$ and outgoing particles $p_1<0$, we find:
\begin{equation}
x(t)=x_0+\frac{\arctanh{[\sin(\lambda t)]}}{\lambda}\, .\label{traj-circ}
\end{equation}

This is the affine null geodesic of AdS spacetime.

\subsection{Undeformed Translations}\label{subsec:class-trans}

Using the canonical brackets \eqref{can-brack}, we can construct the infinitesimal translations in time and space, and the boost by the action of the generators $E_{\circ}$ \eqref{e-circ}, $P_{\circ}$ \eqref{p-circ}, and $N_{\circ}$ \eqref{n-circ}:
\begin{align}
    &\{t,E_{\circ}\}=\cosh{(\lambda x)}\, ,\qquad \{t,P_{\circ}\}=0\, ,\label{t-e-circ}\\
    &\{t,N_{\circ}\}=\frac{\sinh{(\lambda x)}}{\lambda}\, , \qquad \{x,E_{\circ}\}=\sinh{(\lambda x)}\tan{(\lambda t)}\, ,\\
    &\{x,P_{\circ}\}=1\, ,\qquad \{x,N_{\circ}\}=\cosh{(\lambda x)}\frac{\tan{(\lambda t)}}{\lambda}\, .\label{x-n-circ}
\end{align}

Similar rules could be found for momenta, but since we will not use them in this letter, we shall not dwell on them here. In any case, endowed with these tools, we have finite transformations $T_{G;a}$ generated by a charge $G$, parametrized by the quantity $a$, acting on a phase space quantity $X=X(t,x,p_0,p_1)$ (which we call an exponentiation):
\begin{equation}
    T_{G;a}\triangleright X=e^{-a G}\triangleright X=\sum_{n=0}^{\infty}\frac{(-a)^n}{n!}\{X,G\}_n\, ,\label{fin-trans}
\end{equation}
where $\{X,G\}$ is recursively defined as:
\begin{equation}
   \{X,G\}_0=X\, ,\qquad \{X,G\}_n=\{\{X,G\}_{n-1},G\}\, .
\end{equation}

This allows us to translate individual events and trajectories in AdS spacetime. In the following, we shall consider the deformed case and derive similar analytical mechanics results using a deformed version of the AdS algebra.


\section{Deformed Anti-de Sitter Kinematics}\label{sec:q-ads}

In this section, we rely mainly on results reported in \cite{Ballesteros:2016bml}, which describes the $\kappa$-AdS algebra as a deformation of the anti-de Sitter algebra by the presence of a parameter with dimensions of energy $\kappa^{-1} \doteq \ell$, which is supposedly proportional to the invariant quantum gravity energy scale. In $1+1$-dimensions, this algebra presents the following Poisson brackets:
\begin{eqnarray}
    \{N,E\} &=& P\, ,\label{poisson-n-e} \\
    \{E,P\} &=& \lambda^2 N\, ,\label{poisson-e-p} \\
    \{N,P\} &\approx& E - \ell \left[E^2 + \frac{P^2}{2} + \frac{\lambda^2}{2}N^2\right]\, ,\label{poisson-n-p}
\end{eqnarray}

where we are considering only first-order terms in $\ell$. Besides that, the deformed Casimir is
\begin{equation}
    C \approx E^2 - P^2 - \lambda^2 N^2 - \ell \left(EP^2 + \lambda^2 E N^2\right)\, .\label{def-c}
\end{equation}

As can be seen, we can treat this formalism as a $\kappa$-deformation of the AdS algebra (which is recovered when $\ell = \kappa^{-1} \rightarrow 0$), or as a $\Lambda$-deformation of the $\kappa$-Poincar\'e algebra \cite{Ballesteros:2021dob} (which is recovered when $\lambda = \sqrt{-\Lambda} \rightarrow 0$).

Considering the former viewpoint, we can also treat the generators $(E,P,N)$ as $\kappa$-deformations of the AdS ones treated in the previous section $(E_{\circ},P_{\circ},N_{\circ})$. At this point, we need to make some simplifications and bear our intuition in similar approaches previously studied in the literature. Due to the isotropy of this approach, we shall assume that the space translation generator is not modified by the $\kappa$-corrections and is given by the anti-de Sitter one considered in the previous section, i.e., $P = P_{\circ}$. We should stress that this same property applies to the $\kappa$-de Sitter algebra considered by other authors, as can be seen in Eq. (19) of \cite{Amelino-Camelia:2012vzf}. As we will see, this is a valid representation of this algebra. The rest of the analysis is done according to the following ansatz:
\begin{eqnarray}
    N &=& N_{\circ} + \ell f(E_{\circ},P_{\circ},N_{\circ})\, ,\label{new-nn} \\
    E &=& E_{\circ} + \ell(\alpha E_{\circ}^2 + \beta E_{\circ} P_{\circ} + \gamma P_{\circ}^2 + \delta \lambda E_{\circ} N_{\circ} \nonumber \\
    && + \epsilon \lambda P_{\circ} N_{\circ} + \phi \lambda^2 N_{\circ}^2)\, ,\label{new-ee} \\
    P &=& P_{\circ}\, ,\label{new-pp}
\end{eqnarray}
where $f$ is a function of the undeformed charges, and $(\alpha, \beta, \gamma, \delta, \epsilon, \phi)$ are dimensionless parameters that shall be fixed a posteriori. We can find the form of the function $f$ by substituting Eqs. \eqref{new-ee} and \eqref{new-pp} in the relation \eqref{poisson-e-p}. Using the undeformed brackets \eqref{poisson-ec-pc}, \eqref{poisson-nc-pc}, \eqref{poisson-nc-ec} in this expression, we find that the function $f$ must have the following form:
\begin{equation}\label{cond-g}
    f = 2\alpha N_{\circ} E_{\circ} + \beta N_{\circ} P_{\circ} + \frac{\delta}{\lambda}(\lambda^2 N_{\circ}^2 + E_{\circ}^2) + \frac{\epsilon}{\lambda} P_{\circ} E_{\circ} + 2\phi N_{\circ} E_{\circ} \, .
\end{equation}

Now, we use this form of $f$ in $N$ \eqref{new-nn} and consider the bracket $\{N,E\}$ in \eqref{poisson-n-e} (where we also use the brackets involving the undeformed generators \eqref{poisson-ec-pc}, \eqref{poisson-nc-pc}, \eqref{poisson-nc-ec}). Isolating the factors that multiply the product of independent charges $E_{\circ}^2,\, N_{\circ}^2,\, P_{\circ}^2,\, N_{\circ} P_{\circ},\, N_{\circ} E_{\circ},\,  E_{\circ} P_{\circ}$, we find the following set of conditions:
\begin{eqnarray}
    \beta = \delta = 0\, ,\qquad \phi = -2\alpha - \gamma\, ,
\end{eqnarray}
and $\epsilon$ remains an arbitrary parameter. We consider this simplification in the definition of $N$ and $E$, given by \eqref{new-nn} and \eqref{cond-g}, to analyze the bracket $\{N,P\}$. Therefore, using \eqref{poisson-ec-pc}, \eqref{poisson-nc-pc}, \eqref{poisson-nc-ec} into \eqref{poisson-n-p}, we isolate the factors that multiply the products of undeformed charges to further constrain the remaining parameters as:
\begin{eqnarray}
    &\alpha = \beta = \delta = 0\, ,\qquad \gamma = \frac{1}{2} \Rightarrow \phi = -\frac{1}{2}\, , \\
    &\epsilon = \text{arbitrary}\, .
\end{eqnarray}

This leads us to the following expression of the deformed charges in terms of the undeformed ones:
\begin{eqnarray}
    E &=& E_{\circ} + \ell \left(\frac{P_{\circ}^2}{2} - \frac{\lambda^2}{2} N_{\circ}^2 + \epsilon \lambda P_{\circ} N_{\circ}\right)\, ,\label{new-e} \\
    N &=& N_{\circ} + \ell \left(\frac{\epsilon}{\lambda} P_{\circ} E_{\circ} - N_{\circ} E_{\circ}\right)\, ,\label{new-n} \\
    P &=& P_{\circ}\, .\label{new-p}
\end{eqnarray}

To check the consistency of this result, it is sufficient to verify that Eqs. \eqref{poisson-n-e}, \eqref{poisson-e-p}, \eqref{poisson-n-p} are satisfied (using the undeformed expressions \eqref{poisson-ec-pc}, \eqref{poisson-nc-pc}, \eqref{poisson-nc-ec}). From Eq.\eqref{new-n}, we see that to recover the $\kappa$-Poincar\'e algebra in the limit $\lambda\rightarrow 0$, we need to require that $\epsilon=0$. This fixes the ambiguity of this formalism. We can express these charges in cosmological coordinates $(t,x,p_0,p_1)$ by using Eqs. \eqref{e-circ}, \eqref{p-circ}, \eqref{n-circ}. 

Endowed with these expressions, we can calculate the deformed Casimir operator in terms of the undeformed charges as well. Substituting Eqs. \eqref{new-e}, \eqref{new-n}, \eqref{new-p} into the Casimir charge \eqref{def-c}, we find:
\begin{equation}
    C = E_{\circ}^2 - P_{\circ}^2 - \lambda^2 N_{\circ}^2 + {\cal O}(\ell^2) = C_{\circ} + {\cal O}(\ell^2)\, .
\end{equation}

Surprisingly, we do not find first-order corrections on the Casimir operator when we express it in terms of the undeformed charges (and this is a coordinate-independent statement). For this reason, we would find, for example, that the dispersion relation of particles in this scenario is given by the undeformed expression \eqref{casimir}. This implies that the trajectory of a massless particle is unmodified and is given by \eqref{traj-circ}. A similar issue on the difference between the symmetry generators and the energy and momentum of photons is also discussed in footnote 4 of \cite{Carmona:2022pro}.
\par
As we shall see, despite the absence of a deformed trajectory, we will be able to find a time delay at first order in $\ell$ due to the presence of deformed translations generated by $E$ and $P$. In fact, we can express the infinitesimal transformation generated by these charges by using their dependence with the undeformed ones and the actions already calculated in \eqref{t-e-circ}-\eqref{x-n-circ}.


\section{Anti-de Sitter Kinematics in Conformal Coordinates}\label{sec:trans-conf}

Some lessons considered in previous investigations by other authors on the de Sitter case can be applied to ours. For example, the use of conformally flat coordinates significantly simplifies the computations needed to find finite translations. The AdS line element can be cast in conformally flat coordinates by the definition of a conformal time $\eta$:
\begin{equation}
    d\eta = \sec{(\lambda t)} \, dt \Rightarrow \eta = \frac{\arctanh{[\sin(\lambda t)]}}{\lambda}\, ,\label{transf-coord}
\end{equation}
in which the line element \eqref{line-element} assumes the form $ds^2 = \sech^2{(\lambda \eta)}(d\eta^2 - dx^2)$. As the undeformed charges \eqref{e-circ}-\eqref{n-circ} are scalars from the point of view of coordinate transformations, to express them in coordinates $(\eta, x)$, we must simply perform the transformation \eqref{transf-coord} on the time coordinate and correspondingly express the momenta $p_{\mu}$ in the new coordinates (which gives the same result as transforming $K^{\mu}$ to new coordinates and calculating $K'{}^{\mu}p'_{\mu}$). Let us denote the energy and momentum in these new coordinates as $(\Omega, \Pi)$, respectively. One can easily verify that the following relation is valid:
\begin{eqnarray}
    p_0 &=& \frac{d\eta}{dt} \, \Omega = \cosh{(\lambda \eta)} \, \Omega\, , \\
    p_1 &=& \Pi\, .
\end{eqnarray}
For this reason, the undeformed conserved charges read:
\begin{eqnarray}
    \Eo &=& \cosh{(\lambda \eta)} \cosh{(\lambda x)} \, \Omega + \sinh{(\lambda \eta)} \sinh{(\lambda x)} \, \Pi\, ,\label{conf-eo} \\
    \Po &=& \Pi\, ,\label{conf-po} \\
    \No &=& \frac{\cosh{(\lambda \eta)} \sinh{(\lambda x)} \, \Omega}{\lambda} + \frac{\sinh{(\lambda \eta)} \cosh{(\lambda x)} \, \Pi}{\lambda}\, .\label{conf-no}
\end{eqnarray}

In fact, using this representation of the charges, the Casimir \eqref{def-c} takes the form:
\begin{equation}
    C_{\circ} = \cosh^2{(\lambda \eta)} (\Omega^2 - \Pi^2)\, ,
\end{equation}
as expected. Besides that, as the analysis done in the previous section regarding the $\kappa$-AdS algebra was performed in a coordinate-invariant way, the same results discussed previously are preserved in these conformally flat coordinates.


\section{Time delay in $\kappa$-Anti-de Sitter Space}\label{sec:k-time-delay}

In order to connect distant events by the emission and detection of light rays, we need to consider these events as if they were local to relatively distant observers, for instance, Alice and Bob. We shall consider a thought experiment in which massless particles with different energies are emitted in a single event that coincides with the origin of the coordinate system associated with the observer Alice and are detected on the $x$-axis of the observer Bob with a delay in time coordinates.
\par
To find this delay, we need to translate the worldlines from Alice's coordinates to Bob's coordinates, which include translating the initial conditions from Alice's coordinate system to Bob's. A difference in the location of the emission event will appear in Bob's coordinates due to deformed translations, and this is intimately related to the principle of relative locality, in which the location in spacetime of a distant event is described by a distant observer as if it were nonlocal. This is just an apparent nonlocality, since for an observer close enough to the emission event (Alice), it takes place objectively at a single spacetime event.
\par
For this reason, we will need to translate the origin of Alice's coordinate system to express it in terms of Bob's coordinates. This is done by acting with a finite translation as defined in \eqref{fin-trans} as\footnote{We define the translation in spacetime as a translation in space followed by a time translation. This is a choice that eases the derivation computationally. It is sufficient for the purposes of this letter, but other choices could be explored in the future.}
\begin{eqnarray}
    \text{Bob}&=&e^{-a_{\eta} E} \triangleright e^{-a_h P} \triangleright \text{Alice} \Rightarrow\\
    (\eta_0^B,x_0^B)&=&e^{-a_{\eta} E} \triangleright e^{-a_h P} \triangleright (0,0)\, .
\end{eqnarray}

In this representation, the coordinates $(\eta, x, \Omega, \Pi)$ are canonical since the charges are scalar quantities and the Poisson brackets are invariant under coordinate transformations. From the definition of $E$ and $P$ in \eqref{new-e}, \eqref{new-p}, along with the undeformed charges in conformally flat coordinates \eqref{conf-eo}, \eqref{conf-po}, \eqref{conf-no}, we notice that space translations generated by $P$ are undeformed, do not affect the time coordinate, and simply shift the spatial one, while time translations present an energy-dependence. Performing the series of operations defined in \eqref{fin-trans}, we exponentiate the translations as 
\begin{align}
&\eta^B = e^{-a_{\eta} E} \triangleright e^{-a_{x} P} \triangleright \eta^A = e^{-a_{\eta} E} \triangleright\left[\sum_{n=0}^{\infty}\frac{(-a_{x})^n}{n!}\{\eta^A, P\}_n\right]\\
&= \sum_{m=0}^{\infty}\frac{(-a_{\eta})^m}{m!}\left\{\sum_{n=0}^{\infty}\frac{(-a_{x})^n}{n!}\{\eta^A, P\}_n, E\right\}_m\, ,\label{poisson1}\\
&x^B = e^{-a_{\eta} E} \triangleright e^{-a_{x} P} \triangleright x^A = e^{-a_{\eta} E} \triangleright\left[\sum_{n=0}^{\infty}\frac{(-a_{x})^n}{n!}\{x^A, P\}_n\right]\\
&= \sum_{m=0}^{\infty}\frac{(-a_{\eta})^m}{m!}\left\{\sum_{n=0}^{\infty}\frac{(-a_{x})^n}{n!}\{x^A, P\}_n, E\right\}_m\, .\label{poisson2}
\end{align}

Where we use the fact that the new charges are functions of the undeformed ones, and we apply these on the space and time coordinates to derive properties of the deformed finite translations.

We assume that the clock synchronization and determination of distance between Alice and Bob are done using soft, massless particles, i.e., using standard special relativity. Therefore, there is a relation between the translation parameters.

For this reason, Eqs. \eqref{poisson1} and \eqref{poisson2} are actually power series in one of the parameters, for instance, $a_{\eta}$. After a tedious computation, we find the following result when applied to the point $(0,0)$ (confirmed up to 7th order perturbation in $a_{\eta}$)
\begin{eqnarray}
\eta_0^B &=& -\lambda^{-1}\arctanh\left[\sin \left(\lambda a_{\eta}\right)\right] + {\cal O}(\ell^2)\, ,\label{origin-bob-eta}\\
    x_0^B &=& -a_x - \frac{\ell \Pi}{\lambda} \sin{(\lambda a_{\eta})} + {\cal O}(\ell^2)\, ,\label{origin-bob-x}
\end{eqnarray}
where we see $\eta_0^B \big{|}_{\ell=0} = -\lambda^{-1} \arctanh \left[\sin \left(\lambda a_{\eta}\right)\right]$ as the undeformed translated time of Alice's origin, in Bob's coordinates, in standard anti-de Sitter space. This quantity is equal to $a_x$ such that the origins of the two frames are also connected by the trajectory of the massless soft particle. These are the central equations of this letter, and from them, the time delay is calculated straightforwardly.
\par
In fact, in this thought experiment, two massless particles (that we simply call photons) are emitted simultaneously from Alice's origin: one soft photon with momentum $\Pi_s$, whose kinematics is unaffected by $\kappa$-corrections, and one hard photon with momentum $\Pi_h$, whose kinematics presents Planck-scale corrections. As we calculated that both particles present the same dispersion relation at first order in $\ell$ (in terms of undeformed generators), their trajectories are the same concerning the soft and hard photons. Since Alice and Bob's origins are connected by the soft photon trajectory, we set $\eta_0^B|_{\ell=0} = -a_{x}$. Below, we express the photon trajectories in the conformally flat patch using Alice's and Bob's coordinates, where we use the superscript ``$A$'' and ``$B$'' to represent them:
\begin{eqnarray}
x^{A,B}_{s} &=& x^{A,B}_{0;s} + \eta^{A,B}_{s} - \eta^{A,B}_{0;s}\, ,\label{traj-s}\\
x^{A,B}_{h} &=& x^{A,B}_{0;h} + \eta^{A,B}_{h} - \eta^{A,B}_{0;h} + {\cal O}(\ell^2)\, .\label{traj-h}
\end{eqnarray}

The time delay is a difference in the time coordinates of both particles when they reach the same space location in Bob's coordinates, i.e., when $x^{B}_{s} = x^{B}_{h}$ in \eqref{traj-s} and \eqref{traj-h}. We can assume, without loss of generality, that the soft particle reaches the origin of Bob's coordinate system, i.e., $\eta^B_{s} = 0$ in \eqref{traj-s}. Using these conditions, along with \eqref{origin-bob-eta} and \eqref{origin-bob-x}, we subtract Eqs. \eqref{traj-s} and \eqref{traj-h} to find:
\begin{equation}
    \Delta \eta^B = \eta^{B}_{h} - \eta^{B}_{s} = x^{B}_{0;s} - x^{B}_{0;h} = \frac{\ell}{\lambda} \Pi_h \sin{(\lambda a_{\eta})}\, .
\end{equation}

\begin{figure}
    \centering
    \includegraphics[scale=0.5]{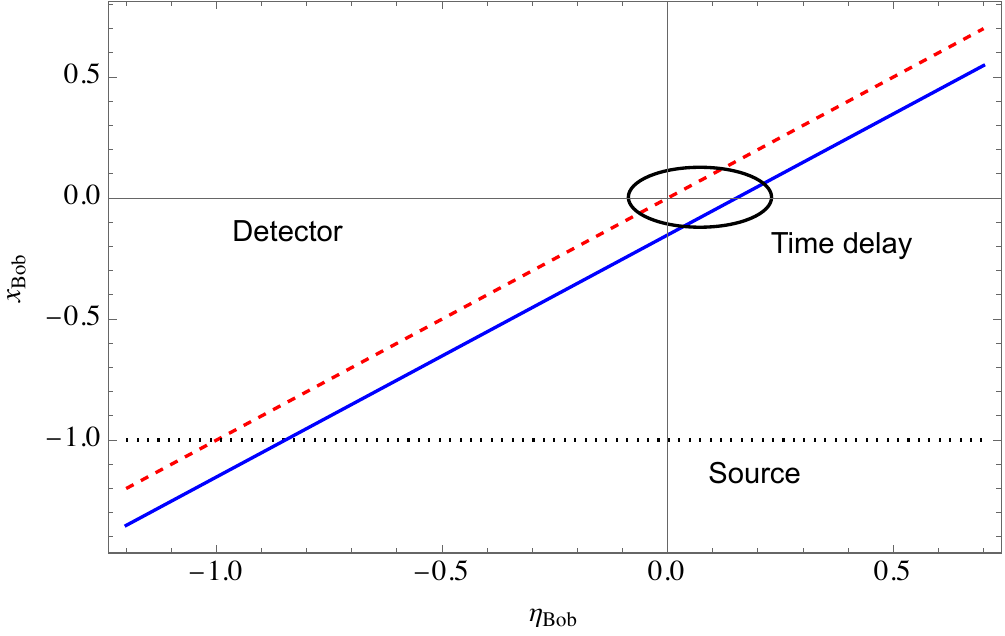}
    \caption{$\kappa$-AdS trajectories of photons in Bob's conformal coordinates. The red/dashed line describes the soft photon emitted at $x^B_{0;s}=-a_x=\eta_0^B\big{|}_{\ell=0}=-1$ in Bob's coordinates. Although the emission occurs at the same spacetime location in Alice's coordinates $(t^A_{0;s},x^A_{0;s})=(t^A_{0;h},x^A_{0;h})=(0,0)$, it seems displaced in Bob's coordinates due to relative locality. The blue/solid line describes the trajectory of the hard photon when $|p|=1$, $\ell=-0.2$, $\lambda=1$, in Bob's coordinates.}
    \label{fig:time-delay-conformal}
\end{figure}

We notice that we have an oscillatory behavior due to the sine function. We depict the relative locality effect in conformal coordinates in Fig.~\ref{fig:time-delay-conformal}. As can be seen, the emission event (local in Alice's coordinate) appears nonlocal in Bob's coordinates. This is the origin of the time delay effect even with undeformed trajectories. In solid red, we have the undeformed emission, propagation, and detection of the soft photon. The dashed black line gives the time delay.

Due to its oscillatory behavior, the time delay can be positive or negative, depending on the distance between the detector and the source. This oscillatory behavior does not exactly align with the intuition of the time delays literature, in which the effect should grow with the distance between the source and the detector. However, the point is that the trajectories in this case are undeformed at first order, so we depend exclusively on the symmetry properties of the anti-de Sitter space, which, as we know, presents exotic behaviors like the presence of closed timelike curves and that some geodesics usually converge, expand, and reconverge. Therefore, such oscillatory behavior should not be a complete surprise. These general properties are discussed in section 5.2 of \cite{Hawking:1973uf}.

In order to compare this result with the Jacob-Piran ansatz, we express it in terms of the redshift function and the cosmological time. By Eq.~\eqref{transf-coord}, we have that the relation between the redshift $z$, cosmological scale factor $a(t)$, and conformal time is the following:
\begin{equation}
    \frac{1}{1+z}=a(t)=\cos{(\lambda t)}=\sech{(\lambda \eta)}\, .
\end{equation}

As the emission occurred at the coordinate $\eta_0^B\big{|}_{\ell=0}$, we have that $\cosh{\lambda \eta_0^B\big{|}_{\ell=0}}=1+z$. Using Eq.~\eqref{origin-bob-eta} and the fact that the hyperbolic cosine is an even function, the translation parameter is related to the redshift as $a_{\eta}=\lambda^{-1} \arcsin\left[\sqrt{z\left(z+2)\right)}/\left(z+1\right)\right]$.

Considering that $t=\lambda^{-1}\arcsin{(\tanh{(\lambda \eta)})}$ and that, as we discussed below Eq.~\eqref{dxdt}, we have $\Pi=p_1\doteq p<0$, we express the time delay in cosmological coordinates as (we see that $\Delta \eta=\Delta t+{\cal O}(\ell^2)$)
\begin{equation}
    \Delta t=-\ell\frac{|p|}{\lambda}\frac{\sqrt{z\left(z+2\right)}}{z+1}\, .\label{k-ads-time-delay}
\end{equation}
This is the main result of our letter. It is the description of a time delay from the $\kappa$-anti-de Sitter algebra. It does not present deformed trajectories, yet due to the presence of deformed translations, a time delay is presented.

For small redshifts, this expression assumes the form
\begin{equation}
    \Delta t|_{z\ll 1}\approx -\frac{\ell |p|}{\lambda}\left[\sqrt{2z}-\frac{3z^{3/2}}{2\sqrt{2}}\right]\, .\label{k-ads-time-delay-small}
\end{equation}

This is the same kind of powers in redshift that would appear had one considered the Jacob-Piran formula \cite{Jacob:2008bw} for an anti-de Sitter space with scale factor $a(t)=\cos{(t)}$ and a modified dispersion relation of the kind $p_0^2-\sec^2{(\lambda t)}p^2+\ell\sec^3{(\lambda t)}|p|^3=0$ (where $H(t)=a(t)^{-1}da(t)/dt$)
\begin{equation}
    \Delta t_{JP,z\ll 1}=\ell |p|\int_0^z\frac{1+\bar{z}}{H(\bar{z})}d\bar{z}\approx -\frac{\ell |p|}{\lambda}\left(\sqrt{2z}-\frac{z^{3/2}}{2\sqrt{2}}\right) \, .\label{JP-time-delay}
\end{equation}

\begin{figure}
    \centering
    \includegraphics[scale=0.55]{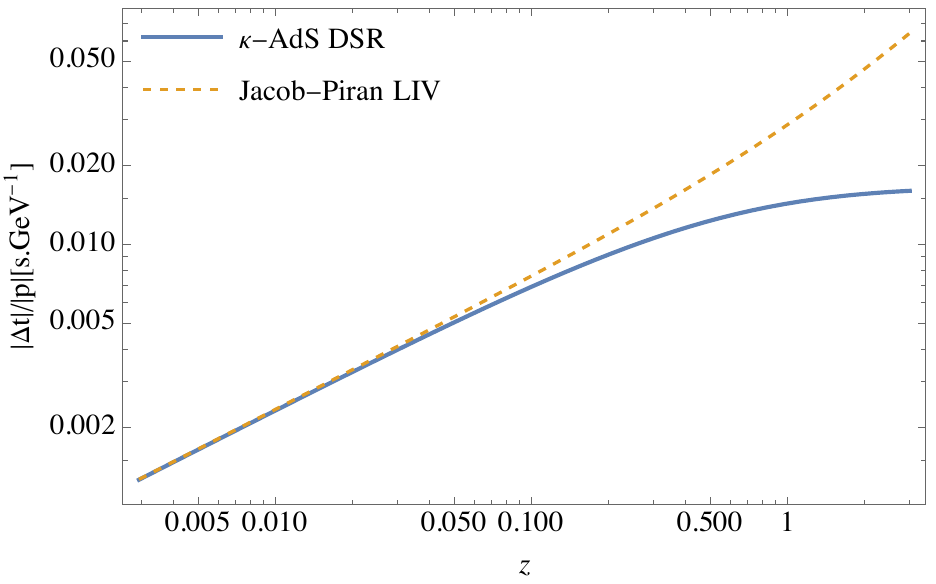}
    \caption{Ratio between the time delay in cosmological coordinates and the hard photon's momentum, as a function of the redshift, for $\ell^{-1}=-E_{\text{Planck}}=-1.2\times 10^{19}\, \text{GeV}$, $\lambda=5\times 10^{-18}\, \text{s}^{-1}$. The blue/solid line describes the $\kappa$-AdS time delay, without MDR and just with relative locality effects \eqref{k-ads-time-delay}. While the orange/dashed line describes Jacob-Piran formula \eqref{JP-time-delay} with an MDR.}
    \label{fig:time-delay2}
\end{figure}

In Fig.~\ref{fig:time-delay2}, we depict the Jacob-Piran time delay of Eq.~\eqref{JP-time-delay} in blue/solid, and the $\kappa$-AdS time delay of Eq.~\eqref{k-ads-time-delay}. We see that both formulas agree for small delays, but then, the deformed translation of the $\kappa$-AdS algebra interferes in the time delay. The delay that we presented is bounded to be $\Delta t|_{z\rightarrow \infty}=-\ell \lambda^{-1}p$, despite the relative distance between the events. Other exotic behaviors can be found in recent analyses carried out in \cite{Amelino-Camelia:2023srg}, where a richer class of delays is presented.

The accelerated expansion of the universe implies that for small redshifts we can describe the spacetime geometry by a de Sitter metric, and the $\kappa$-de Sitter algebra could be used to constrain the quantum gravity parameter at the dark energy-dominated epoch. The main point of this letter is the construction of a time delay from the deformation of translations in a class of quantum algebras that was not previously explored to derive this observable. We expect that in the future we can use these results as a starting point to construct a more realistic scenario that combines pieces of anti-de Sitter propagation to construct an FLRW-based time delay that is compatible with the DSR principles, similar to what was done in \cite{Rosati:2015pga} for the $\kappa$-de Sitter case.


\section{Final Remarks}\label{sec:conc}

We found the relation between the generators of the $\kappa$-anti-de Sitter algebra in $1+1$ dimensions \cite{Ballesteros:2016bml} and the undeformed anti-de Sitter algebra at first order in the quantum gravity parameter. In the basis used in \cite{Ballesteros:2016bml}, we found that the boost and spacetime translations are deformed. However, we discovered that there is no deformation of the Casimir charge when we express it in terms of the undeformed quantities. This means that the trajectories are undeformed at the first order of correction.

In a Lorentz invariance-violating scenario, the presence of undeformed trajectories implies the absence of a time delay when particles with different energies are emitted simultaneously and detected on Earth \cite{Addazi:2021xuf}. However, if we assume that the relativity principle is preserved, i.e., that the local frames of the emission and detection events are related by a deformed translation, then a remnant first-order effect remains. This result has been derived using the $\kappa$-de Sitter algebra in the past \cite{Amelino-Camelia:2012vzf}, but for the first time, we perform a derivation using the $\kappa$-anti-de Sitter algebra.

A novel result is the emergence of a bounded time delay, which is related to the behavior of the deformed translations in anti-de Sitter spacetime. AdS spacetime has found a wide range of applicability from cosmology \cite{Magueijo:2009ff} to other areas of physics \cite{Gibbons:2011sg}, which deserves to be explored within this context in future works.

Nevertheless, the $\kappa$-de Sitter results have been used as the cornerstone for the derivation of a time delay formula that would be compatible with local relativistic principles, through the creative construction of a propagation of signals and communication frames in a quantum FLRW spacetime by infinitesimal delays based on the $\kappa$-de Sitter algebra with different curvatures \cite{Rosati:2015pga}. Since there is no privileged role of positive curvature in comparison to the negative one from geometric and algebraic perspectives, we wonder what kind of result would be found if one used the $\kappa$-AdS algebra as the fundamental bricks to build such a delay in the effective FLRW spacetime.

The main challenge of this approach is how to handle the composition of coordinates of the $\kappa$-AdS algebra, which is given by trigonometric functions, while in the $\kappa$-dS algebra, the cosmological and conformal coordinates furnish computationally simpler functions. Despite these difficulties, a first and necessary step towards this analysis is the derivation of such a delay for a single curvature, which was carried out in this letter. In the future, we aim to extend the results of this letter on two fronts: the exploration of different bases of the $\kappa$-AdS algebra in order to derive more general results that include deformed trajectories; and the construction of the effective FLRW time delay based on the $\kappa$-AdS algebra building blocks.

\section*{Acknowledgments}
I. P. L. was partially supported by the National Council for Scientific and Technological Development - CNPq grant 312547/2023-4 and by the grant 3197/2021, Para\'iba State Research Foundation (FAPESQ).



\bibliographystyle{elsarticle-harv} 
\bibliography{adstd}

\end{document}